\documentclass[journal]{IEEEtran}

\let\labelindent\relax
\usepackage{amsmath,amssymb}
\usepackage{algorithmic,algorithm}
\usepackage{bm}
\usepackage{epsfig}
\usepackage{graphicx}
\usepackage{subfigure}
\usepackage{float}
\usepackage{cite}
\usepackage{url}
\usepackage{color}
\usepackage{balance}
\usepackage{mdwlist}
\usepackage{multirow}
\usepackage{threeparttable}
\usepackage{enumitem}
\usepackage{amsmath}
\usepackage{stmaryrd}
\usepackage{booktabs}
\usepackage{siunitx}
\usepackage{multirow}
\usepackage{threeparttable}
\usepackage{graphicx}
\usepackage{epsfig,amsmath,amsfonts,cite}
\newcommand{\thickhline}{\noalign{\hrule height 1.0pt}}
\DeclareMathAlphabet\mathbfcal{OMS}{cmsy}{b}{n}
\newcommand{\ten}[1]{\mathbfcal{#1}}
\newcommand{\mat}[1]{\mathbf{#1}}

\newtheorem{definition}{Definition}

\def\sssp{\def\baselinestretch{0.88}\large\normalsize}\sssp

\begin{document}

\title{Sparse Tucker Tensor Decomposition on a Hybrid FPGA-CPU Platform}

\author{Weiyun Jiang, \and Kaiqi Zhang, \and Colin Yu Lin, \and Feng Xing, \and Zheng~Zhang,~\IEEEmembership{Member,~IEEE}
\thanks{This work was supported by NSF Award 1817037.}
\thanks{W. Jiang, K. Zhang and Z. Zhang are with the the Department of Electrical and Computer Engineering, University of California, Santa Barbara, CA 93106, USA (e-mail:  {weiyunjiang@ucsb.edu, kzhang70@ucsb.edu, and zhengzhang@ece.ucsb.edu).}}
\thanks{C. Y. Lin and F. Xing are with Xilinx Inc., Beijing, China (e-mail:  {YULIN1@xilinx.com and FENGX@xilinx.com).}}}

\maketitle

 \markboth{\MakeLowercase{Accepted by}
 IEEE TRANSACTIONS ON COMPUTER-AIDED DESIGN OF INTEGRATED CIRCUITS AND SYSTEMS, ~Vol. ~XX, No.~XX,~XX~2020}{Jiang \MakeLowercase{\textit{et al.}}:Title}

\begin{abstract}


Recommendation systems, social network analysis, medical imaging, and data mining often involve processing sparse high-dimensional data. Such high-dimensional data are naturally represented as tensors, and they cannot be efficiently processed by conventional matrix or vector computations. Sparse Tucker decomposition is an important algorithm for compressing and analyzing these sparse high-dimensional data sets. When energy efficiency and data privacy are major concerns, hardware accelerators on resource-constraint platforms become crucial for the deployment of tensor algorithms. In this work, we propose a hybrid computing framework containing CPU and FPGA to accelerate sparse Tucker factorization. This algorithm has three main modules: tensor-times-matrix (TTM), Kronecker products, and QR decomposition with column pivoting (QRP). In addition, we accelerate the former two modules on a Xilinx FPGA and the latter one on a CPU. Our hybrid platform achieves $23.6 \times \sim 1091\times$ speedup and over $93.519\% \sim 99.514 \%$  energy savings compared with CPU on the synthetic and real-world datasets.


\end{abstract}

%
%
%
%


\section{Introduction}
As massive data is collected from social media, wearable devices and internet of things, novel algorithms and platforms are highly desired to handle data-intensive computing tasks. Vector- and matrix-based methods can efficiently process $1$-way data (e.g., a sequence of voice data) or $2$-way  data (e.g., a gray-scale image), but they are often inefficient to handle multi-way data. Representative examples includes $3$-way (or order-3) E-commerce data (which records customers' preference on massive products over a few months), $4$-way (or order-$4$) cardiac image data (which records the spatial data of $3$D at multiple time points). Processing such multi-way data often suffers from the curse of dimensionality. 

Tensors are a high-order generalization of matrices and vectors, and they are a natural tool to represent and process multi-way data~\cite{kolda2009tensor}. Leveraging various tensor decomposition or factorization  methods~\cite{kolda2009tensor,oseledets2010tt,de2000multilinear,de2000best}, the curse of dimensionality of storing and computing multi-way data can be avoided or significantly mitigated in many applications. For instance, the canonical polyadic (CP)~\cite{candecomp,harshman1970fpp} and tensor-train~\cite{oseledets2010tt} factorizations can reduce the storage cost and unknown variables from an exponential function to a linear one. Tucker factorization~\cite{de2000multilinear} can be used for high-order principle component analysis or facial recognition~\cite{vasilescu2002multilinear,vasilescu2003multilinear,vasilescu2005multilinear}. Tensor computation has achieved tremeonduous success in data mining~\cite{kolda2008scalable}, computer vision~\cite{vasilescu2002multilinear,vasilescu2003multilinear,vasilescu2005multilinear}, medical imaging~\cite{batmanghelich2011regularized}, electronic design automation~\cite{zhang2016big,zhang2014enabling,zhang2016tensor,luan2019prediction} and deep learning~\cite{novikov2015tensorizing,ttrnn2017icml,hawkins2019bayesian}.

The emerging tensor computation concept brings in massive research opportunities and challenges on the hardware level.  Due to the fundamental difference between tensor and matrix computations, we may need to re-think many aspects of tensor computation (e.g., storage, computing and data movement) on specific platforms. Increasing research results have been reported to improve the tensor data storage and computing on the cloud and high-performance clusters ~\cite{kaya2016high,smith2017sparse,li2015input}. However, little work has been done on resource-constrained platforms. This becomes increasingly important as the need of energy-efficient machine learning and data privacy surges. In order to address this issues, some efforts have been made towards tensor-compressed neural networks on mobile devices~\cite{kim2015compression} and dense tensor operations on FPGA. For instance, some dense tensor operations including MTTKRP, TTM and TTMc were accelerated in~\cite{srivastava2019t2s}; a spectral analysis of Hankel tensors was reported in \cite{huang2019high}. To perform dense Tucker decomposition on FPGA, Zhang et al.~\cite{zhang2019tucker} divided the hardware architectures into three modules: tensor-times-matrix, singular value decomposition via Jacobi iterations and tensor permutation/reshaping. In addition, a warm-start algorithm  was used to reduce the cost of Jacobi iterations. The resulting FPGA accelerator demonstrated significant speedup compared with both CPU and GPU. However, the FPGA accelerator~\cite{zhang2019tucker} cannot exploit data sparsity, and it becomes energy- and time-inefficient when dealing with sparse tensors. Ref.~\cite{srivastava2020tensaurus} reported some sparse tensor computation kernels. For instance, it demonstrated how to implement both dense and sparse tensor operations, such as sparse TTMc via sparse compute pattern $SF^3$. To our best knowledge, there is no FPGA accelerator available for sparse Tucker decomposition. 

In this paper, we investigate the hardware acceleration of Tucker factorization for {\bf sparse} tensor data. Sparse tensors widely appear in practice due to the missing information in recommendation systems, medical image or E-commerce data. For instance, in magnetic resonance imaging (MRI), one can generate a sparse tensor by partial MRI scanning, then reconstruct the whole image with a low cost~\cite{roohi2016dynamic}. In neuroscience, researchers use sparse tensors to monitor the brain variability~\cite{fillard2005extrapolation}. In EDA, it is often too expensive to obtain all simulation or measurement data, thus one uses a partially sampled sparse tensor for process variation or performance uncertainty prediction~\cite{zhang2016big,zhang2016tensor,luan2019prediction}. Although extensive algorithms have been developed to process sparse tensors, their hardware/algorithm co-optimization remains a rarely explored field~\cite{zhang2019tucker}. This task has become increasingly important as energy efficiency and privacy cause lots of concerns in the data science and machine learning community.

\begin{figure*}[t]
	\centering
		\includegraphics[width=4.5in]{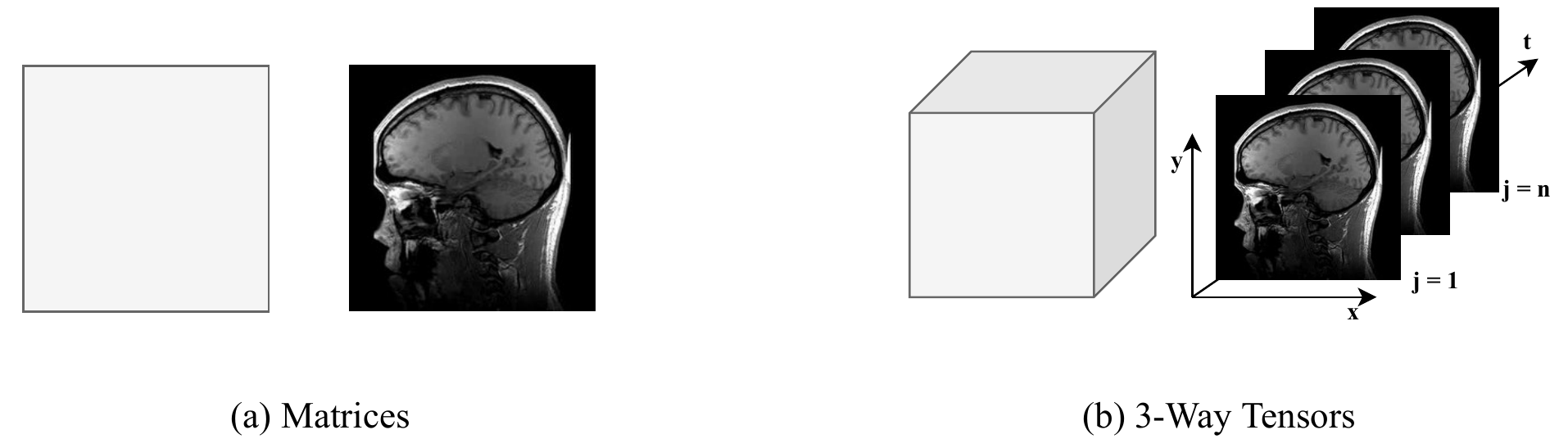} 
\caption{(a) A matrix is a $2$-D data array (e.g., one slice of MRI data), (b) a $3$-way tensor is a $3$-D data array (e.g., multiple slices of images).}
	\label{fig:tensor}
\end{figure*}

\subsection{Paper Contributions and Organization}
This paper proposes to design an energy- and memory-efficient hybrid FPGA-CPU accelerator for sparse Tucker decomposition~\cite{kaya2016high}. This algorithm consists of three major components: tensor-times-matrix (TTM)~\cite{kolda2009tensor}, Kronecker product~\cite{van2000ubiquitous} and QR decomposition with column pivoting (QRP)~\cite{golub1996matrix}. Our specific contributions include:
\begin{itemize}
\item On the hardware side, we present a high-level synthesis (HLS) FPGA implementation for sparse Tucker decomposition. We describe the design of two modules, TTM and Kronecker product, by exploiting the data sparsity. 
\item On the algorithm side, we replace the conventional singular value decomposition (SVD)~\cite{golub1971singular} with the QR decomposition with column pivoting (QRP)~\cite{golub1996matrix} to reduce the data storage cost and to speed up the computation.
\item We implement our FPGA accelerator in a Xilinx FPGA on Amazon web service (AWS). Then we compare our hybrid FPGA-CPU accelerator with CPU and with the recently developed dense FPGA accelerator~\cite{zhang2019tucker} on synthetic and real-world sparse tensor benchmarks. Our hybrid FPGA-CPU accelerator
achieves $1.15 \times$$ \sim$$ 1091\times$ speedup and consumes $93.519\% $$\sim$$ 99.514 \%$ less energy. In addition, our proposed accelerator achieves significant speedup ($23.6\times$$\sim $$ 1091\times$) when the tensor is very large and sparse
\end{itemize}

This paper is organized as follows. Section~\ref{tensorref} introduces some background information about tensor operations. Section~\ref{sec:Sparse_tucker} presents the algorithm and our Vivado HLS FPGA design of a sparse Tucker decomposition. We compare our FPGA/CPU hybrid platform with CPU and the dense Tucker FPGA accelerator \cite{zhang2019tucker} in terms of run-time and energy efficiency in Section~\ref{sec:Experiments}. Finally, Section~\ref{sec:conclusion} concludes this paper.



\section{Preliminaries of Tensors}
\label{tensorref}
This section presents some background about tensors, which is necessary for understanding the ideas of this paper. 


\begin{definition}
A  tensor $\ten{X} \in \mathbb{R}^{I_1 \times I_2 \times \dots \times I_N}$ is a high-dimensional array of order $N$. Here the order $N$ (also known as ``way") is the total number of dimensions. A matrix $\mat{X} \in \mathbb{R}^{n_1\times n_2}$ is a $2$nd-order (or $2$-D) tensor, and its element indexed by $(i_1, i_2)$ can be denoted as $x_{i_1i_2}$. For a general $N$th-order (or $N$-way) tensor $\ten{X}$, its element indexed by $(i_1, i_2 \cdots, i_N)$ is denoted as $x_{i_1 i_2\cdots i_N}$.  
\end{definition}

Fig.~\ref{fig:tensor} shows a matrix (e.g., one slice of MRI data) and  a $3$-way tensor, respectively.  In this paper, we use boldface lower-case letters (e.g., $\mat{x}$) to denote vectors, boldface upper-case letters (e.g., $\mat{X}$) to denote matrices, and boldface Euler script letters ( e.g.,$\ten{X}$) to denote tensors. A scalar is denoted by a lower-case letter, e.g., $x$.





\begin{definition}
The inner product of two tensors with the same size is defined as
\begin{equation}
\langle \ten{X}, \ten{Y} \rangle =\sum \limits_{i_1i_2 \cdots i_N} x_{i_1i_2 \cdots i_N} y_{i_1i_2 \cdots i_N}.
\end{equation}
Furthermore, the Frobenius norm (also known as F-norm) of a tensor $\ten{X}$ is defined as $|| \ten{X}||_{\rm F} =\sqrt{\langle \ten{X}, \ten{X} \rangle}$.
\end{definition}
\begin{definition}
A matricization operation, (also known as unfolding or flattening), reshapes a tensor into a matrix. The mode-$n$ matricization of a tensor $\ten{X} \in \mathbb{R}^{I_1 \times I_2 \times \dots \times I_N}$ is denoted as $\mathbf{X}_{(n)}$ which has $I_n$ rows and $\prod_{k\neq n}I_k$ columns. Element-wise, we have each entry of $\mathbf{X}_{(n)}$ as
\begin{equation}
    \mathbf{X}_{(n)}({i_n, j})=x_{i_1i_2\cdots i_N} \\
    {\text{with}}~j = 1 + \sum_{k=1, k \neq n}^{N} (i_k-1)\prod_{m=1, m\neq n}^{k-1} I_m.
\end{equation}
\end{definition}
\begin{definition}
The mode-$\mathbf{n}$ tensor matrix product [or tensor-times-matrix (TTM)], between a tensor $\ten{X} \in \mathbb{R}^{I_1 \times I_2 \times \dots \times I_N }$ and a matrix $\mathbf{U} \in \mathbb{R}^{J \times I_n}$ is denoted as 
\begin{equation}
    \ten{G} = \ten{X}\times_n \mathbf{U},~{\text{where}}~{\ten{G} \in \mathbb{R}^{I_1 \times \dots \times I_{n-1} \times J \times I_{n+1}\times \dots \times I_N}}.
\end{equation} 
Element-wise, we can write this operation as
\begin{equation}
    g_{i_1\dotsi_{n-1} j i_{n+1}\dots i_N} = \sum_{i_n=1}^{I_n} x_{i_1 i_2 \dots i_N}u_{j i_n}.
\end{equation}
We may also obtain a TTM product by using the unfolded tensors:
\begin{equation}
    \ten{G} = \ten{X} \times_n \mathbf{U} \Leftrightarrow \mat{G}_{(n)} = \mat{U}\mat{X}_{(n)}.
\end{equation}
\end{definition}



We further introduce a matrix operation that will be used in our subsequent tensor computation.
\begin{definition}
Given a matrix $\mat{A} \in \mathbb{R}^{m\times n}$ and another  matrix $\mat{B} \in \mathbb{R}^{p\times q}$, their Kronecker product $\mat{A}\otimes\mat{B}$ is the following matrix $\mat{C} \in \mathbb{R}^{mp\times nq}$
\begin{equation}
\mat{C}=\mat{A}\otimes\mat{B}=\begin{bmatrix}
 a_{11}\mat{B} & \cdots & a_{1n}\mat{B}\\ 
 \vdots & \ddots & \vdots\\ 
 a_{m1}\mat{B} & \cdots & a_{mn}\mat{B}
\end{bmatrix}.
\end{equation}
\end{definition}

\begin{algorithm}[t]
\caption{Standard HOOI for Tucker Decomposition}
\label{alg:hooi}
\begin{algorithmic}[1]
\STATE { Initialize $\{\mat{U}_n\} _{k=1}^N$ via HOSVD}
\WHILE {not converge} 
    \FOR {$n=1,2, \ldots, N$} 
        \STATE {$\ten{Y} = \ten{X} \times_1 \mat{U}_1^T \dots \times_{n-1} \mat{U}_{n-1}^T \times_{n+1} \mat{U}_{n+1}^T \dots \times_N \mat{U}_N^T$}
        \STATE {Unfold $\ten{Y}$ and perform SVD: $ \mat{Y}_{(n)}=\mat{U} \mat{S} \mat{V}^T $}
        \STATE {$\mat{U}_n \leftarrow$ the first $R_n$ columns of $\mat{U}$.}
    \ENDFOR
\ENDWHILE
\STATE {\textbf{return} $\{ \mat{U}_{n}\}_{n=1}^N$. }

\end{algorithmic}
\end{algorithm}

\section{Accelerator for Sparse Tucker Decomposition}
\label{sec:Sparse_tucker}
Given a tensor $\ten{X} \in \mathbb{R}^{I_1 \times I_2 \times \dots \times I_N}$, the Tucker decomposition~\cite{de2000best} approximates it with a small low-rank core tensor $\ten{G} \in \mathbb{R}^{R_1 \times R_2 \times \dots \times R_N}$ and $N$ factor matrices $\{ \mat{U}_n \in \mathbb{R}^{I_n \times R_n } \}_{n=1}^N$:
\begin{equation}
\ten{X}\approx \ten{G}\times_1 \mat{U}_1 \times_2 \mat{U}_2 \cdots \times_N \mat{U}_N.
\end{equation}
Here $(R_1, R_2, \cdots, R_N)$ is a multilinear tensor rank.

The Tucker decomposition can be regarded as a high-order generalization of singular value decomposition (SVD), and it is often implemented with the power iteration method called high-order orthogonal iteration (HOOI) in~\cite{de2000best}. As shown in Alg.~\ref{alg:hooi}, it aims to find the orthogonal matrices $\{ \mat{U}_n \in \mathbb{R}^{I_n \times R_n }\}_{n=1}^N$ to maximize the F-norm of 
\begin{equation}
\label{TTM_G}
    \ten{G} = \ten{X}\times_1\mathbf{U}_1^T \times_2 \mathbf{U}_2^T \dots \times_N \mathbf{U}_N^T.
\end{equation}
In every iteration, we need to compute the $R_n$ dominant left singular vectors of unfolded matrix $\mat Y_{(n)}$, where
\begin{equation}
\label{eqn:power_ite}
    \ten{Y} = \ten{X} \times_1 \mathbf{U}_1^T \dots \times_{n-1} \mathbf{U}_{n-1}^T \times_{n+1} \mathbf{U}_{n+1}^T \dots \times_N \mathbf{U}_N^T.
\end{equation}
The orthogonal matrix is obtained by a SVD of the unfolded matrix $\mat Y_{(n)}$. 

The standard HOOI becomes very inefficient for sparse tensors because Line 4 of Alg.~\ref{alg:hooi} does not exploit any data sparsity and always performs $N-1$ times of TTM operations. 

\begin{algorithm}[t]
\caption{Sparse Tucker Decomposition}
\label{alg:sparse_tucker}
\textbf{Input: } 
A sparse tensor $\ten X$
                \\$R_1$,$\ldots$,$R_N$: rank of approximation

\begin{algorithmic}[1]
\STATE \textbf{initialize} $\mathbf{U_1},...,\mathbf{U_N}$ randomly.
\REPEAT
\FOR {$n=1, 2, \dots, N$}
\FOR {$x_{i_1,\dots,i_N} \neq 0$}
\STATE $\mathbf{Y}_{(n)}(i_n,:) \mathrel{+}= x_{i_1,\dots,i_N}[\otimes_{t\neq n}\mathbf{U}_t(i_t,:)]$
\ENDFOR
\STATE $\mathbf{U}_n \leftarrow \textbf{QRP}(\mathbf{Y}_{(n)}, R_n)$
\ENDFOR
\STATE $\ten{G} \leftarrow \ten{Y}\times_N \mathbf{U}_N^T$
\UNTIL{convergence or maximum number of iterations reached} 
\end{algorithmic} 
\textbf{Output: }
\\$\ten{G}$: a $R_1 \times $$\ldots$$\times R_N$ core tensor 
\\$\mathbf{U_1},\ldots,\mathbf{U_N}$: $\mat{U}_n$ is a $R_n \times I_n $ factor matrix
\end{algorithm}

\begin{figure}[t]
	\centering
		\includegraphics[width=3.5 in]{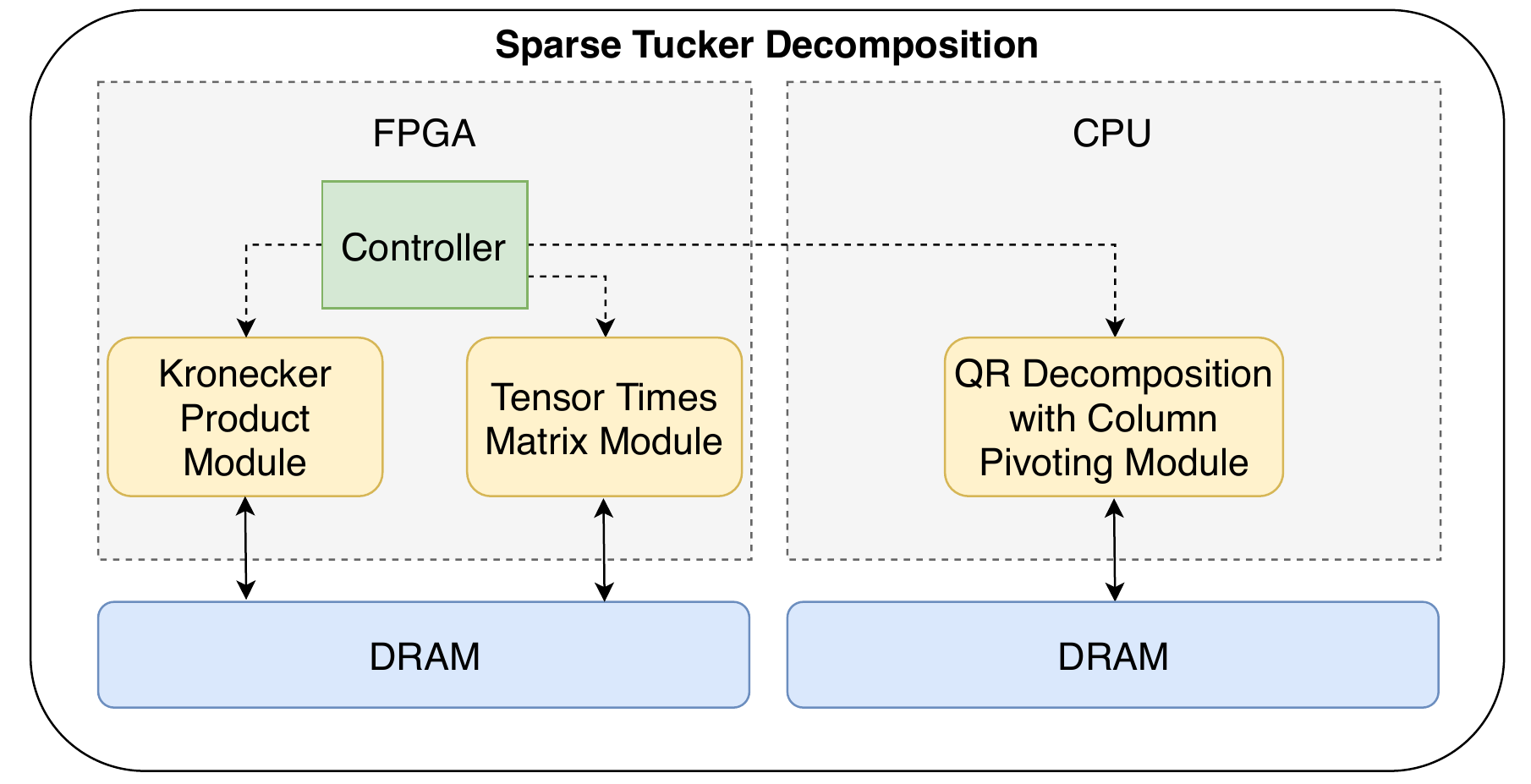}
\caption{A Hybrid FPGA-CPU platform for sparse Tucker factorization.}
	\label{fig:Block_Diag}
\end{figure}

\begin{table}[t]
\caption{Coordinate (COO) format of a $5\times5\times5\times5$ sparse tensor. Here $(i,j,k,l)$ denotes an index, and $nnz$ is the value of an associated non-zero data element.}
\centering
\begin{tabular}{|c|c|c|c|c|}
\hline
$i$ & $j$ & $k$ & $l$ & $nnz$ \\ \hline \thickhline
1 & 1 & 1 & 1 & 2   \\ \hline
1 & 1 & 1 & 5 & 7.5 \\ \hline
1 & 1 & 3 & 5 & 4   \\ \hline
2 & 2 & 2 & 4 & 5   \\ \hline
\end{tabular}
\label{tab:COO}
\end{table}
\subsection{Overall Algorithm Flow}

In this paper, we design an FPGA-CPU hybrid accelerator based on~\cite{kaya2016high} to perform Tucker factorization for sparse tensors. Two formats can be used to represent sparse tensors:
\begin{itemize}
    \item The coordinate format (COO) stores a sparse tensor with all nonzero elements and their associated coordinate vectors, shown in~Table \ref{tab:COO}. The first four columns represent the coordinate $(i,j,k,l) $ of 4 nonzero elements, and the last column represents the corresponding value. The COO format usually requires storage of $O(nnz*N)$ index values and $O(nnz)$ nonzero data values, where $nnz$ is the number of nonzero elements and $N$ is the mode of the tensor.
    \item Compressed sparse fiber format (CSF) stores a sparse tensor by compressing the indices of nonzero elements that share the same coordinates. It is regarded as high dimensional version of the compressed sparse row (CSR) or compressed sparse column (CSC) formats used for matrices in \cite{gustavson1972some}. The CSF format requires $O(2*(nnz+s+f)+2)$ to store an order-$3$
    tensor with $s$ slices, $f$ fibers and $nnz$ non-zero values.
\end{itemize}
 
In this paper, we use the COO format because of its flexibility and simplicity. Furthermore, the COO format provides better performance on merging-related TTM~\cite{tew2016investigation}. If we do not assume any special structure of the tensor and the non-zero elements are uniformly distributed, there will be rarely multiple nonzero elements in a given fiber. In such a general case, the CSF format barely has any advantages in storage compression.

The algorithm flow is summarized in Alg.~\ref{alg:sparse_tucker}. Compared with the standard dense Tucker factorization, the following techniques are used to exploit the data sparsity:
\begin{itemize}
    \item Instead of storing the whole tensor, we only store the nonzero entries by specifying their values and indices. 
    \item When performing the tensor-times-matrix  (TTM) in~\eqref{eqn:power_ite}, we do not perform $N-1$ levels of iterations over all modes except mode $n$. Instead, we only consider the non-zero elements of $\ten{X}$ and have a one-level iteration over the indices of all non-zero elements in $\ten{X}$. 
    \item In order to reduce the computational and memory cost of extracting orthogonal matrix factor $\mat{U}_n$, we replace the SVD of $\mat{Y}_{(n)}$ with a QR decomposition with column pivoting (QRP).
\end{itemize}

The proposed accelerator architecture is shown in Fig.~\ref{fig:Block_Diag}. Because it is difficult to parallelize the QRP operation, we implement it on CPU. Both \eqref{TTM_G} and~\eqref{eqn:power_ite} require TTM operations, but they are handled in different ways. For \eqref{TTM_G} we only need to compute 
\begin{equation}
\label{eqn:core_ten}
    \ten{G} = \ten{Y}\times_N \mat{U}_N
\end{equation}
once for each iteration after obtaining $\ten{Y}$ (which is often dense) by \eqref{eqn:power_ite}. Therefore, we design a specialized TTM module on FPGA in Section~\ref{subsec:TTMc}. For the power iteration in \eqref{eqn:power_ite}, we design a Kronecker product module on FPGA to accelerate the sparse operation, which is detailed in Section~\ref{subsec:kron}.


\subsection{Tensor-Times-Matrix (TTM) on FPGA}
\label{subsec:TTMc}
The computation of $\ten{G}$ in \eqref{TTM_G} requires $N$ tensor-matrix products on the original huge-size tensor $\ten{X}$. This expensive computation actually can be simplified.

Assuming that we have already done the power iteration \eqref{eqn:power_ite} for $n=N$, and obtained a small-size tensor $\ten{Y} \in \mathbb{R}^{R_1\times R_2\times \dots \times I_N}$ and an orthogonal factor matrix $\mathbf{U}_N\in \mathbb{R}^{I_N\times R_N}$. We only need to compute the mode-$N$ tensor-matrix product \eqref{eqn:core_ten} to obtain the core tensor $\ten{G}$ (line $9$, Alg. \ref{alg:sparse_tucker}). This TTM can be written in an element-wise manner:
\begin{equation}
\label{eqn:ttm_ele}
    (\ten{Y}\times_N \mathbf{U}_N^T)_{r_1 r_2\dots r_N} = \sum_{i_N=1}^{I_N} y_{r_1 r_2 \dots i_N} \mat{U}_N (i_N, r_N).
\end{equation}
Equivalently, we can express this particular TTM with unfolded tensors as follows: 
\begin{equation}
\label{eqn:core_unfolded}
     \ten{G} = \ten{Y}\times_N \mat{U}_N^T \Leftrightarrow \mat{G}_{(N)} = \mat{U}_N^T \mat{Y}_{(N)}.  
\end{equation}
Here $\ten{G}_{(N)}$ and $\mat{Y}_{(N)}$ are the mode-$N$ unfolding of the tensors $\ten{G}$ and $\ten{Y}$, respectively. 

\begin{algorithm}[t]
\caption{Vivado HLS Implementation of TTM on $3$-way Tensors}
\label{alg:HLS_ttm}

\begin{algorithmic}[l]
\REQUIRE $\mat{Y} \in \mathbb{R}^{R_1R_2\times I_3}, \mat{U} \in \mathbb{R}^{R_3\times I_3}$
\STATE $\ell = R_1 R_2$, $b = 32$
\FOR {$(i_{b} = 0;i_{b} < \ell;i_{\delta} \mathrel{+}=b)$}
    \STATE  \textbf{initialize} $\textbf{tmp}$ as zero
    \FOR{$(k =0;k<R_3;k\texttt{++})$}
        \FOR{$(i_o = 0;i_o<b;i_o\texttt{++})$}
            \FOR{$(t=0;t<I_3;t\texttt{++})$}
                \STATE $\textbf{tmp}[i_o, k] \mathrel{+}= \mat{Y}[i_o+i_{b}, t]*\mat{U}[k, t]$
            \ENDFOR
        \ENDFOR
    \ENDFOR
    \FOR{$(k = 0;k<R_3;k\texttt{++})$}
        \FOR{$(i_o=0;i_o<b;i_o\texttt{++})$}
            \STATE $\mat{G}[i_o+i_{b}, k] = \textbf{tmp}[i_o, k]$
        \ENDFOR 
    \ENDFOR
\ENDFOR
\end{algorithmic}
\textbf{Output: } $\mat{G} \in \mathbb{R}^{R_1 R_2\times R_3} $
\end{algorithm}

\begin{figure}[t]
	\centering
		\includegraphics[width=3 in]{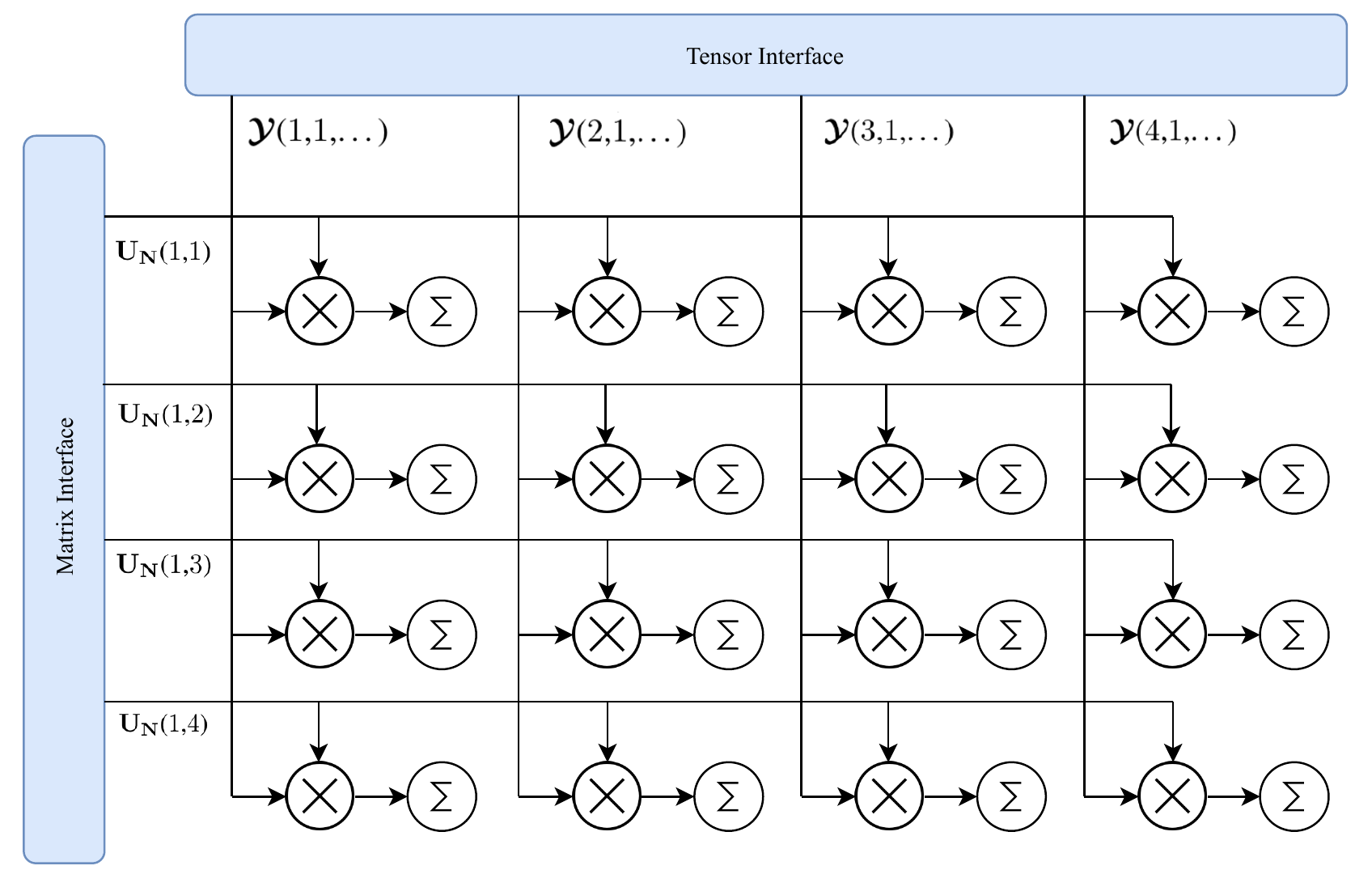}
\caption{Tensor-times-matrix (TTM) data flow.}
	\label{fig:ttm_flow}
\end{figure}
\begin{figure}[t]
	\centering
		\includegraphics[width=1.6 in]{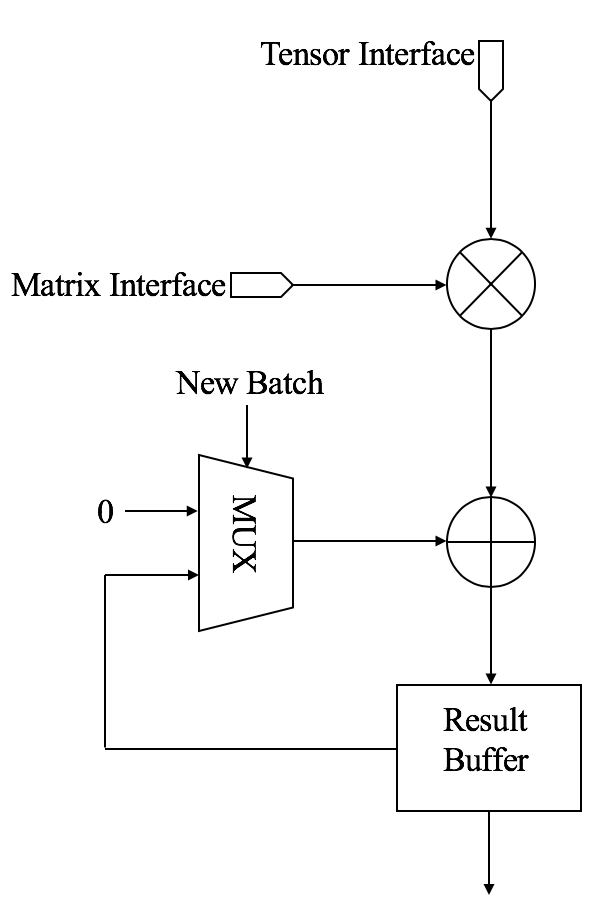}
\caption{Tensor-times-matrix (TTM) Processing Element (PE) \cite{zhang2019tucker}.}
	\label{fig:pe_ttm}
\end{figure}

In FPGA design, the $3$-D sparse tensor $\ten{X} \in \mathbb{R}^{I_1\times I_2 \times I_3}$ is stored with a cost $O(nnz)$, where $nnz$ denotes the number of nonzero elements. However, the tensor $\ten{Y} \in \mathbb{R}^{R_1\times R_2\times I_3}$ in (\ref{eqn:core_ten}) is dense, and we need to store all of its elements. Although $\ten{Y}$ is multi-dimensional, it is unnecessary to create a new copy of this tensor. We can just reshape it into a 2-D matrix of size $R_1 R_2 \times I_3$. 
Meanwhile, it is critical to process the entries of $\ten{Y}$ in several batches. The batch size, $b$, controls the number of entries in $\ten{Y}$, being processed in each iteration. If we set the batch size as $b = R_1R_2$, we will end up with 3 nested for-loops because the outermost for-loop is redundant. As a result, all the entries of $\ten{Y}$ have to be processed at the same time, resulting in an extremely large amount of loop unrolling, which is not practical when $R_1R_2$ is larger. To overcome this issue, we decrease our batch size to $32$, and separate this loop into two parts, resulting in 4 nested for-loops to compute the resultant tensor of the TTM. In this way, we could achieve optimal loop unrolling on memory-constrained FPGAs. 

We provide the Vivado HLS implementation pseudo code of the TTM for a $3$-way tensor $\ten{X}$ in Alg. \ref{alg:HLS_ttm}. Given a $3$-way tensor, $\ten{X} \in \mathbb{R} ^{I_1 \times I_2 \times I_3}$, \eqref{eqn:core_ten} is a mode-$3$ TTM between $\ten{Y} \in \mathbb{R}^{R_1 \times R_2 \times I_3}$ and $\mathbf{U} \in \mathbb{R}^{I_3\times R_3}$, where $\ten{G} \in \mathbb{R} ^{R_1\times R_2\times R_3}$ is the result. In the pseudo code, we reshape our tensors $\ten{Y} \in \mathbb{R}^{R_1 \times R_2 \times I_3}$ and $\ten{G} \in \mathbb{R} ^{R_1\times R_2\times R_3}$ into matrices $\mat{Y} \in \mathbb{R}^{R_1 R_2 \times I_3}$ and $\mat{G} \in \mathbb{R}^{R_1R_2\times R_3}$. Basically, we divide our result, $\mat{G}$, into several portions such that we can update one portion of $\mat{G}$ in each batch:
\begin{itemize}
    \item First, we initialize the temporary matrix, $\mathbf{tmp}$ as zero matrix of size $b \times R_3$, where $b$ is the batch size. This temporary matrix stores one portion of our result $\mat{G}$. 
    \item Then, we compute TTM by multiplying unfolded tensor $\mat{Y}$ and $\mat{U}$ based on (\ref{eqn:core_unfolded}) and store the results in $\mathbf{tmp}$. 
    \item Finally, we just update one portion of $\mat{G}$ with $\mathbf{tmp}$. 
\end{itemize} 
In order to optimize the Vivado HLS implementation, we reshape $\mat{U}$ in cyclic forms by a factor of $8$, and we reshape $\mat{Y}$ and $\mathbf{tmp}$ in cyclic forms by a factor of $16$. Furthermore, in order to save RAM usage, we assign only one port of RAM to the variables, $\mat{Y}$, $\mat{U}$, and $\mathbf{tmp}$. We also assign the intermediate variable $\mathbf{tmp}$ to registers instead of memory to minimize memory usage.

Fig.~\ref{fig:ttm_flow} shows the data flow in the TTM computation module on FPGA. According to the element-wise formula (\ref{eqn:ttm_ele}), each entry of the resultant tensor can be recognized as the sum of product between the entries from the original tensor $\ten{Y}$ and the entries from the matrix $\mathbf{U_N}$. In Fig.~\ref{fig:ttm_flow}, it shows that data from the tensor interface, $y_{r_1r_2\dots i_N}$ multiplies with the data from the matrix interface, $\mathbf{U_N}(i_N,r_N)$. After the multiplication, the results are summed up to obtain the entries in the resultant tensor, $(\ten{Y}\times_N \mathbf{U}_N^T)_{r_1 r_2\dots r_N}$.

A detailed data flow of the PE for TTM is shown in Fig.~\ref{fig:pe_ttm}, which was proposed in \cite{zhang2019tucker}. A buffer temporarily stores the intermediate result after multiplying the tensor and the matrix. For each batch, the multiplexer selects and adds the intermediate result to the new product. Once all batches are processed, the final result is stored the DRAM.

\subsection{Kronecker Products on FPGA}
\label{subsec:kron}
The power iteration \eqref{eqn:power_ite} requires $O(R^d\times n)$ operations, and it consumes most of the computational power and run-time in the sparse Tucker decomposition. Although an FPGA design was presented in~\cite{zhang2019tucker} to accelerate power iterations, existing design cannot handle sparse tensor data efficiently. Therefore, leveraging~\cite{kaya2016high,van2000ubiquitous}, we design an FPGA module to compute the power iteration via Kronecker products. 

We consider a sparse $3$-way tensor $\ten{X}$ as an example. We investigate the power iteration of mode 1, which is written as $\ten{Y}$ = $\ten{X}$ $\times_2$ $\mathbf{U_2}^T$  $\times_3$ $\mathbf{U_3}^T$. To exploit the sparsity, we may choose to compute the Kronecker products and consider only nonzero elements $x_{ijk} \neq 0$~\cite{kaya2016high}:
\begin{equation}
    \mathbf{Y_{(1)}}(i,:) =\mathbf{Y_{(1)}}(i,:)+ x_{ijk}[\mathbf{U_2}(j, :) \otimes \mathbf{U_3}(k, :)].
\end{equation}
The number of Kronecker products depends on the number of nonzero elements in $\ten{X}$, which is often very small for sparse tensors. Furthermore, a Kronecker product can be re-used for all non-zero elements that share the same indices $(j,k)$ for the $2$nd and $3$rd modes. Therefore, replacing TTM of \eqref{eqn:power_ite} with some Kronecker products can largely reduce the computational complexity. Additionally, directly computing TTM is memory-inefficient when the size and order of $\ten{X}$ are large, causing a high cost of RAM and registers on FPGA. 

\begin{algorithm}[t]
\caption{Vivado HLS Implementation of Kronecker Product}
\label{alg:HLS_kron}
\begin{algorithmic}[1]
\STATE {\textbf{Input: } $\mathbf{a} \in \mathbb{R}^{1\times R_2}$, $\mathbf{b} \in \mathbb{R}^{1\times R_3}$ }
\FOR {$(i = 0; i < R_2; i++)$}
    \FOR {$(j = 0; j<R_3;j++)$}
       \STATE $\mathbf{c}[R_3 \times i + j] =\mathbf{a}[i] \times \mathbf{b}[j]$
    \ENDFOR
\ENDFOR
\STATE{\textbf{Output: } $\mathbf{c}\in \mathbb{R}^{1\times R_2R_3}$}
\end{algorithmic}

\end{algorithm}

In the Vivado HLS implementation, we utilize nested for-loops to implement the Kronecker product (Alg. \ref{alg:HLS_kron}):
\begin{itemize}
\item In order to parallelize the Kronecker product on FPGA, we pipeline the first for-loop and unroll the second for-loop. The rank of approximation, $R_1$, $R_2$, and $R_3$, are usually very small compared with the mode sizes. Therefore, the available memory, lookup tables (LUTs) and registers are often sufficient for parallelization. 

\item To update the corresponding rows of unfolded data $\mat{Y}_{(n)}$ in the the power iteration, we simply multiply the Kronecker product result in the LUTs with the corresponding nonzero element $y_{r_1 r_2 \dots i_N}$.
\item In addition, different nonzero elements may share the same index of some modes. In this case, we accumulate the multiplications between these nonzero elements and their corresponding Kronecker product results.  
\end{itemize}

\begin{figure}[t]
	\centering
		\includegraphics[width=3 in]{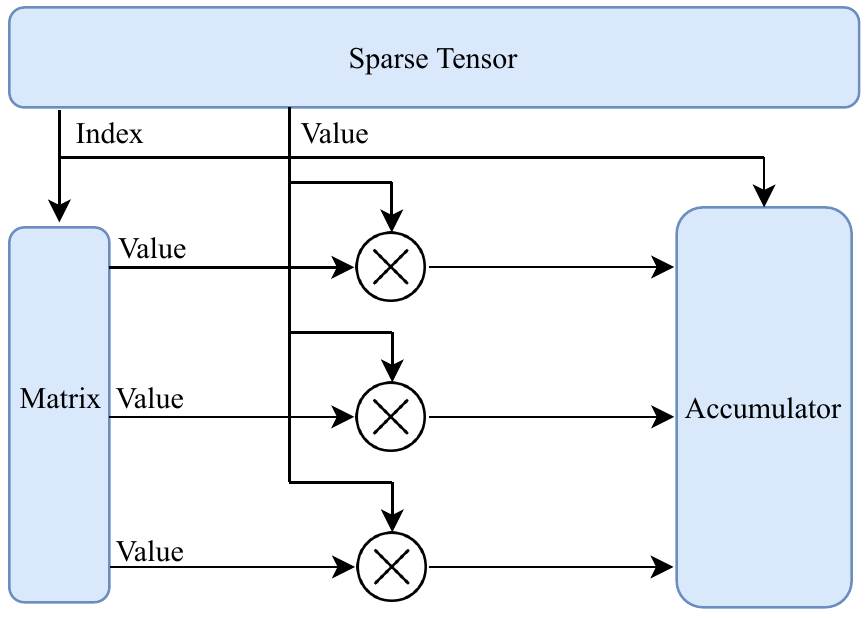}
\caption{The data flow of a Kronecker product.}
	\label{fig:Kronecker_flow}
\end{figure}

Fig.~\ref{fig:Kronecker_flow} shows the data flow inside our Kronecker product module on FPGA. To begin with, the indices of the non-zero elements in the original tensor are extracted. Then, based on the indices of the nonzero entries, the corresponding rows of the orthogonal matrix factor, $\mathbf{U}_t(i_t,:)$ are selected. Assuming there are two row vectors, every entry in one row vector multiply with every entry in the other row vector to generate the Kronecker product. Since we only compute the Kronecker product between two row vectors (not two matrices), the module only requires multiplication units (no addition units).

\begin{table}[t]
\centering
\caption{Accuracy comparison of Tucker decomposition with SVD and with QRP.}
\begin{tabular}{|c|c|c|}
\hline
Tensor Size     & \begin{tabular}[c]{@{}c@{}}Tucker Decomposition \\ with SVD\end{tabular} & \begin{tabular}[c]{@{}c@{}}Tucker Decomposition \\ with QRP\end{tabular} \\ \hline
\thickhline
$50 \times 50 \times 50$    & $1.9222\times 10^{-09} $                                                           & $1.9228\times 10^{-09}$                                                            \\ \hline
$100 \times 100 \times 100$ & $1.3846\times 10^{-09}$                                                       & $1.3820\times 10^{-09}$                                                            \\ \hline
$200 \times 200 \times 200$ & $1.1588\times 10^{-09} $                                                           & $1.1786\times 10^{-09}$                                                            \\ \hline
$400 \times 400 \times 400$ & $1.2114\times 10^{-09}$                                                            & $1.2115\times 10^{-09}$                                                            \\ \hline
$800 \times 800 \times 800$ & $3.8450\times 10^{-10}$                                                            & $3.8531\times 10^{-10}$                                                            \\ \hline
\end{tabular}
\label{tab:svd_qrp}
\end{table}



\begin{table*}[t]
\centering
\caption{Performance comparison of FPGA and CPU on the TTM task.}
\begin{tabular}{|l|l|l|l|l|l|}
\hline
\multicolumn{1}{|c|}{\multirow{2}{*}{Tensor Size}} & \multicolumn{1}{c|}{\multirow{2}{*}{Matrix Size}} & \multicolumn{2}{c|}{CPU}    & \multicolumn{2}{c|}{FPGA}   \\ \cline{3-6} 
\multicolumn{1}{|c|}{}                             & \multicolumn{1}{c|}{}                             & Run-Time & Energy & Run-Time & Energy \\ \thickhline
$32 \times 32 \times 32$                           & $32 \times 32$                                    & $0.493$ ms       & $22.19$ mJ    & $0.148$ ms      & $0.4212$ mJ   \\ \hline
$32 \times 32 \times 64$                           & $32 \times 64$                                    & $0.596$ ms      & $26.82$ mJ   & $0.281$ ms      & $0.8000$ mJ   \\ \hline
$32 \times 32 \times 128$                          & $32 \times 128$                                   & $1.165$ ms      & $52.43$ mJ    & $0.546$ ms      & $1.556$ mJ    \\ \hline
$32 \times 32 \times 256$                          & $32 \times 256$                                   & $2.021$ ms      & $90.95$ mJ    & $1.077$ ms       & $3.067$ mJ    \\ \hline
\end{tabular}\normalsize
\label{table:ttm_whole}
\end{table*}

\begin{table*}[t]
\centering
\caption{Performance comparison  of FPGA and CPU on the Kronecker product task.}
\begin{tabular}{|l|l|l|l|l|l|}
\hline
\multicolumn{1}{|c|}{\multirow{2}{*}{Size of $\mat{x}_j$}} & \multicolumn{1}{c|}{\multirow{2}{*}{Size of $\mat{x}_k$}} & \multicolumn{2}{c|}{CPU}        & \multicolumn{2}{c|}{FPGA}           \\ \cline{3-6} 
\multicolumn{1}{|c|}{}                                     & \multicolumn{1}{c|}{}                                     & Run-Time  & Energy & Run-Time & Energy  \\ \thickhline
$1 \times 32$                                              & $1 \times 32$                                             & $9.655$ $\mu$s          & $0.4345$ mJ    & $0.578$ $\mu$s          & $2.111$ $\mu$J        \\ \hline
$1 \times 64$                                              & $1 \times 64$                                             & $14.72$    $\mu$s      & $0.6624$ mJ   & $2.301$ $\mu$s          & $8.403$ $\mu$J        \\ \hline
$1 \times 128$                                             & $1 \times 128$                                            & $24.87$ $\mu$s         & $1.119$ mJ    & $9.195$ $\mu$s          & $33.58$ $\mu$J        \\ \hline
$1 \times 256$                                             & $1 \times 256$                                            & $48.24$  $\mu$s        & $2.171$ mJ    & $38.55$ $\mu$s         & $140.7$ $\mu$J        \\ \hline
\end{tabular}\normalsize
\label{table:kron_whole}
\end{table*}

\subsection{QR Decomposition with Column Pivoting}
\label{subsec:QRP}
In existing dense and sparse Tucker factorization~\cite{de2000best,kaya2016high}, the orthogonal matrix $\mat{U}_n$ is obtained with a singular value decomposition (SVD)~\cite{golub1971singular} of $\mat{Y}_{(n)}$. The SVD is accurate but extremely slow at computing the orthogonal matrices. In order to speed up the computation and minimize the memory usage, we propose to use QR decomposition with column pivoting (QRP)~\cite{golub1996matrix} to obtain $\mat{U}_n$. The QRP implementation does not lose any accuracy compared with the SVD implementation. This is clearly shown in Table~\ref{tab:svd_qrp}, which reports the errors of several low-rank Tucker decomposition with both SVD and QRP implementations, respectively. 

Given a matrix $\mathbf{A}\in \mathbb{R}^{m\times n}$, the QRP get an orthogonal matrix $\mathbf{Q}\in \mathbb{R}^{m\times n}$ and an upper-triangular matrix $\mathbf{R}\in \mathbb{R}^{n \times n}$:
\begin{equation}
    \mat{AP} = \mat{QR},
\end{equation}
with $\mat{P}$ being a permutation matrix. 
The $\mathbf{P}$ is chosen so that the diagonal elements of $\mathbf{R}$ is non-increasing:
\begin{equation}
    \mid r_{11}\mid  \geq \mid r_{22}\mid \geq \dots \geq \mid r_{nn}\mid.
\end{equation}
A QRP costs about $2mn^2-2n^3/3$ flops, and an SVD costs about $ 2mn^2 + 11n^3$ flops, where $m \geq n$. In the sparse Tucker factorization of a tensor $\ten{X} \in \mathbb{R}^{I_1 \times I_2 \times \dots \times I_N }$, $\mat{A}$ is $\mat{Y}_{(n)}$, the mode-$n$ unfolding of the tensor $\ten{Y}$ obtained in \eqref{eqn:power_ite}. Consequently, $m=I_n$, $n=\prod \limits_{k\neq n} R_n$, and the computational saving is huge when the tensor order $N$ or multilinear rank parameters $(R_1, R_2, \cdots, R_N)$ are large. In some particular cases, we may end up with a fat rectangular matrix, $\mathbf{Y}_{(n)}$ ($n > m$). In this case, we can perform QRP on a square matrix, $\mathbf{Y}_{(n)}\mathbf{Y}_{(n)}^T$. 

{\bf QRP Implementation.} The QRP in our implementation is based on the Householder reflection. This method computes the orthogonal matrix $\mathbf{Q}$ as the product of multiple Householder reflection matrices:
\begin{equation}
\label{eqn:house_reflect}
    \mat{Q} = \mat{H_1}\mat{H_2}\dots \mat{H_{m-2}}\mat{H_{m-1}}.
\end{equation}
The $j$-th reflection matrix, $\mathbf{H}_j$, is defined as:
\begin{equation}
    \mathbf{H}_j = \mathbf{I} - 2\mathbf{v}_j \mat{v}_j^T = \mathbf{I} - 2\frac{\mat{u}_j \mat{u}_j^T}{\mat{u}_j^T\mat{u}_j},
\end{equation}
 where $\mathbf{u}_j$ is an unit vector and $\mathbf{u}_j = \frac{\mat{v}_j}{\left\lVert \mat{v}_j \right\rVert}$. Vector $\mat{v}_j$ can be chosen based the $j$th column of $\mat{A}$, $\mathbf{a}_j$:
\begin{equation}
    \mathbf{v}_j = \mathbf{a}_j + {\text{sign}}(a_{jj})\left\lVert \mat{a}_n\right\rVert \mat{e}_1. 
\end{equation}
 During every iteration of QRP, we need to update $\mathbf{A}$ by multiplying it with the Householder matrix $\mathbf{H}$. In order to generate the permutation matrix, $\mathbf{P}$, we need to compare the norms of the columns of the updated matrix $\mathbf{A}$ at every iteration, arranging the columns so that the norms of the columns are in descending order. In this way, we can place the most weighted entries in the upper left corner of $\mathbf{Q}$, achieving the similar accuracy to SVD. Since we need to compare the norms of the columns at each iteration, the QRP operation is sequential. In other words, the comparison of the column norms made it very difficult to parallelize the algorithm on FPGA. Thus, we implement the Householder QR decomposition~\cite{golub1996matrix} with column pivoting on CPU.

\section{Results}
\label{sec:Experiments}
This section shows the performance of our hybrid FPGA-CPU accelerator on both synthetic and real-world datasets. We first verify the performance of individual FPGA modules for the TTM and Kronecker product. Afterwards, we verify the performance of the whole FPGA-CPU sparse Tucker accelerator and compare it with CPU. We use the FPGA model XCVU9P-FLGA2577-3-e in our experiment. The maximum frequency of the FPGA implementation is $890$MHz. The CPU model used is Intel(R) Core(TM) i7-6820HK CPU @ 2.70GHz. The size of the RAM is $16$GB. The CPU has a maximum memory bandwith of 
34.1 GB/s and a thermal design power (TDP) of $45$W. In the experiments, we prioritize the computations on CPU to achieve the maximum performance, therefore, the energy consumption on CPU can be estimated as the product of runtime and TDP. We estimate the energy cost of sparse Tucker decomposition on FPGA on Xilinx Vivado via Amazon Web Service. The communication protocol between FPGA and CPU is PCI express, which has a maximum bandwidth of 10GB/s. Our design can also be implemented on a low-end FPGA such as Zynq-7100 as well. On a low-end FPGA, We may decrease the LUT utilization by adjusting the unroll factor in our TTM module implementation.

\subsection{Performance of Individual FPGA Modules}
\label{sec:result}
Firstly we verify the performance of the TTM and Kronecker-Product modules on some synthetic tensor data, and summarize their performance below:

\begin{itemize}
    \item {\bf TTM Module:} We verify the performance by considering a set of 3-way tensors $\ten{Y} \in \mathbb{R}^{R_1\times R_2\times I_3}$ and factor matrices $\mathbf{U} \in \mathbb{R}^{R_3 \times I_3} $. The rank of approximate, $R_1$, $R_2$ and $R_3$, are always very small compared with the original tensor size for data compression. Thus, we set $R_1 = R_2 = R_3 = 32$. The original tensor size, $I_3$ is set to increase from $32$ to $256$ as shown in Table~\ref{table:ttm_whole}. In the real-life examples, the original tensor size $I_3$ can definitely be larger than $256$. And the performance of the tensor-times-matrix (TTM) module won't perform significantly worse when the original tensor size becomes extremely large. Here, we set the maximum of our tensor size to be $256$ for experimental purpose only. The FPGA achieves $1.560\times$ to $3.331\times$ speedup than CPU on these tensor-matrix products. We also compare the energy consumption between FPGA and CPU on the tensor-times-matrix task. As shown in Table~\ref{table:ttm_whole}, the FPGA saves $95.6\%$ to $98.1 \%$ of energy compared with CPU.
    
    \item {\bf Kronecker-Product Module:} As shown in Section 4.3, the Kronecker product used in the sparse Tucker decomposition deals with two row vectors, $\mat{x}_j \in \mathbb{R}^{1 \times R_j}$ and $\mat{x}_k\in \mathbb{R}^{1 \times R_k}$. Therefore, we compare the performance of Kronecker products on FPGA and CPU by changing the rank parameters $R_1$ and $R_2$ from $32$ to $256$. The rank of approximation $R_1$ and $R_2$ does not necessarily need to be equal to each other. We set $R_1$ and $R_2$ to be equal for experimental purpose only. In addition, the rank of approximation $R_1$, $R_2$ and $R_3$ are usually very small compared with the original tensor size for data compression. We increase the rank from $32$ to $256$ to demonstrate the performance of Kronecker product module. We estimated the power of the CPU to be $45$W. The energy consumption of CPU is estimated by multiplying the power with the CPU time. The results are shown in Table~\ref{table:kron_whole}. The speedup of FPGA over CPU ranges from $1.251\times$ to $16.704\times$. As shown in Table~\ref{table:kron_whole}, FPGA consumes $93.519\%$ to $99.514\%$ less energy than CPU on the Kronecker-product tasks. 

\end{itemize}

\begin{figure}[t]
	\centering
		\includegraphics[width=3.5in]{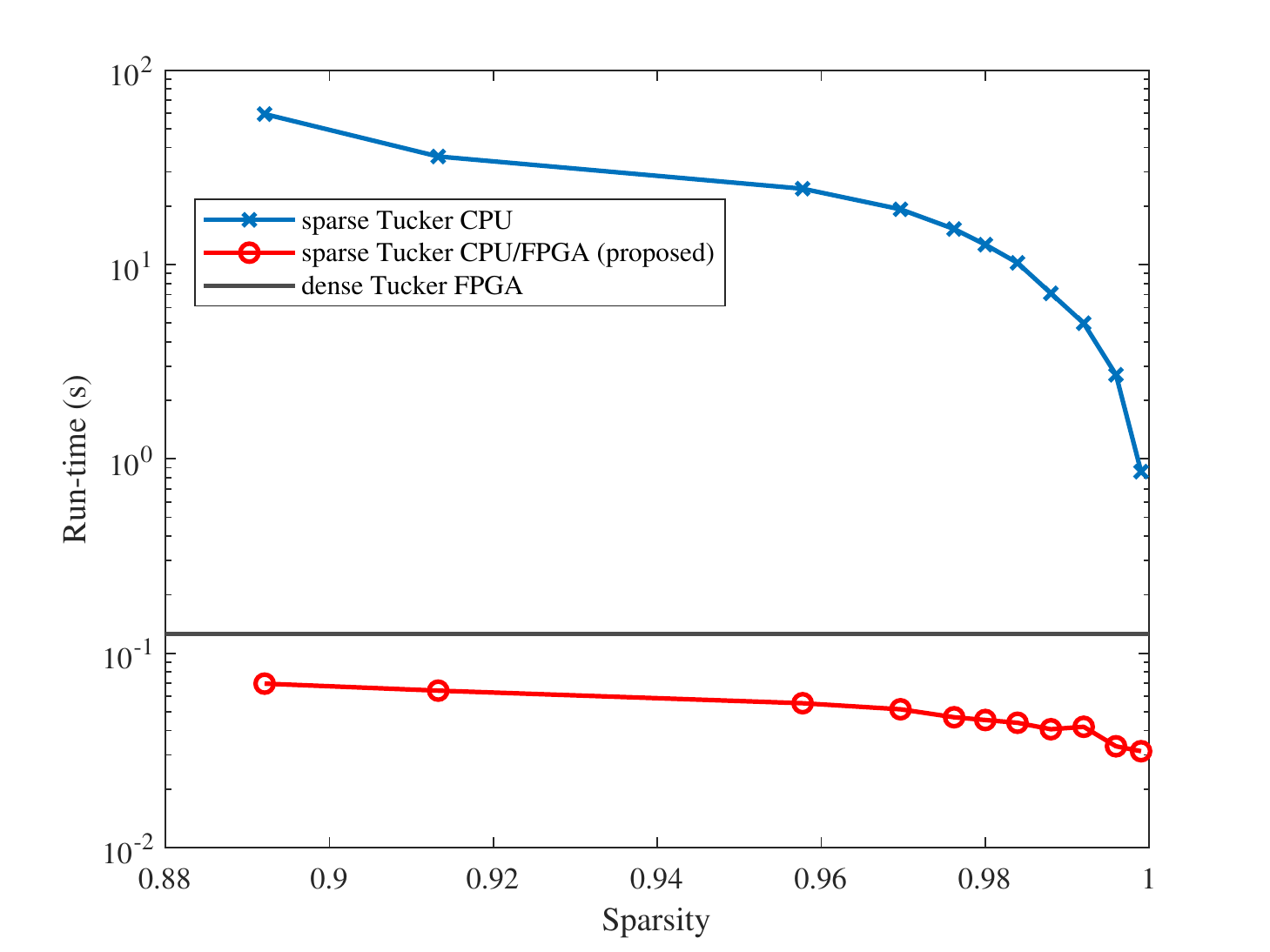} 
\caption{Run-time comparison between the proposed hybrid platform, dense FPGA accelerator and CPU on a set of $200\times 200\times 200$ synthetic random tensors with different sparsity.}
	\label{fig:sparsity_time}
\end{figure}

  \subsection{Accelerator's Performance: Synthetic Datasets}
 
Now we evaluate the whole hybrid FPGA-CPU accelerator on some randomly generated synthetic sparse tensor data sets. Specifically, we consider a set of $3$-way tensors $\ten {X} \in \mathbb{R}^{200 \times 200 \times 200}$ with different sparsity. We fix the rank parameters $R_1$=$ R_2$=$R_3$=$16$. 

Fig. \ref{fig:sparsity_time} compares the run-time of our hybrid FPGA-CPU platform with CPU and densor FPGA accelerator~\cite{zhang2019tucker}. The speedup of the hybrid FPGA-CPU accelerator is $27\times\sim853\times$ compared with CPU. The speedup of our sparse Tucker accelerator is $1.167\times\sim126\times$ faster than the FPGA accelerator designed for dense Tucker decompsition~\cite{zhang2019tucker}. In the whole sparse Tucker decomposition algorithm, the Kronecker product module takes the most amount of time. However, this module is parallelized in our design, and it is significantly sped up on FPGA as shown in Section \ref{sec:result}. When the tensor has more non-zero elements, more Kronecker-product operations are required, leading to a more significant speedup on FPGA.





\begin{table*}[]
\centering
\caption{Performance of Sparse Tucker Decomposition on real-world benchmarks.}
\begin{tabular}{|l|l|c|c|c|c|}
\hline
\multicolumn{2}{|l|}{Benchmarks}                       & Amazon                          & Nell-2                        & Parallel Matrix Multiplication   & Retinal Angiogram                \\ \thickhline
\multicolumn{2}{|l|}{Tensor Size}                      & $20K \times 20K \times 20K $                & $1K \times 1K \times 1K$                  & $25 \times 25 \times 25 $                    & $130 \times 150$                        \\ \hline
\multicolumn{2}{|l|}{Sparsity}                         & $1.128 \times 10^{-10}$ & $2.40 \times 10^{-5}$ & $8\times10^{-3}$         & $0.18$                             \\ \hline
\multirow{2}{*}{CPU}                        & Run-Time & $100.045$ s                       & $7.355$ s                       & $8.175\times10^{-2}$ s   & $0.1838$ s                         \\ \cline{2-6} 
                                            & Energy   & $4502.03$ J                       & $330.98$ J                      & $3.68$ J                           & $8.27$ J                           \\ \hline
{Hybrid FPGA/CPU} & Run-Time & $86.785$ s                        & $0.403$ s                       & $2.179 \times 10^{-3}$ s & $9.898 \times 10^{-3}$ s \\ \cline{2-6} 
     (proposed)                                        & Energy   & $3896.08$ J                       & $17.10$ J                       & $0.1057$ J                         & $0.4667$ J                         \\ \hline
Dense FPGA Tucker~\cite{zhang2019tucker}                                 & Run-Time & $9.47 \times 10^4$ s  & $9.5$ s                         & $9.9 \times 10^{-3}$ s   & $1.18 \times 10^{-2}$ s  \\ \hline
\end{tabular}
\label{tab:realExample}
\end{table*}

\begin{table*}[]
\centering
\caption{Utilization of FPGA on real-world benchmarks. In the column of "memory" we list the number of BRAM, where each BRAM has $18\times 10^3$ bits.}
\begin{tabular}{|c|c|c|c|c|c|c|c|c|c|}
\hline
\multicolumn{2}{|c|}{Name}                               & Expression & Instance & Memory & Multiplexer & Register & Total  & Available & Utilization (\%) \\ \thickhline
\multirow{4}{*}{Amazon}                         &   BRAM\_18K & -          & -     & $542$ & -           &                                                 -        & $542$    & $4320 $      & $13$                                                              \\ \cline{2-10} 
                                                & DSP48E & -          & $282$  & -    & -           & -        & $282$    & $6840$      & $4$                \\ \cline{2-10} 
                                                & FF     & 0          & $17257$ & -   & -           & $107670$   & $124927$ & $2364480$   & $5$                \\ \cline{2-10} 
                                                & LUT    & $406251$     & $17649$    & -& $20587$       & -        & $443268$ & $1182240$   & $37$               \\ \hline
\multirow{4}{*}{Nell-2}                         &   BRAM\_18K & -          & -     & $63$ & -           &                                                      -        & $63$    & $4320 $      & $1$                                                          \\ \cline{2-10} 
                                                & DSP48E & -          & $470$  & -    & -           & -        & $470$    & $6840$      & $7$                \\ \cline{2-10} 
                                                & FF     & $0$          & $29495$   & - & -           & $54691$    & $84186$  & $2364480$   & $4$                \\ \cline{2-10} 
                                                & LUT    & $405656$     & $30863$    & - & $13972$       & -        & $450491$ & $1182240$   & $38$               \\ \hline
\multirow{4}{*}{\begin{tabular}[c]{@{}c@{}}Parallel\\ Matrix \\ Multiplication\end{tabular}}                                   &   BRAM\_18K & -          & -     & $2$ & -           &                                                     -        & $2$    & $4320 $      & $\sim0$                                                            \\ \cline{2-10} 
                                               & DSP48E & -          & $16$       & -           & -       & -& $16$     & $6840$      & $\sim0$          \\ \cline{2-10} 
                                                & FF     & $0$          & $759$ &-     & -           & $107$      & $866$    & $2364480$   & $\sim0$          \\ \cline{2-10} 
                                                & LUT    & $49799$      & $778$ & -     & $707$         & -        & $51284$  & $1182240$   & $4$                \\ \hline
\multirow{4}{*}{\begin{tabular}[c]{@{}c@{}}Retinal \\ Angiogram\end{tabular}}                                                    &   BRAM\_18K & -          & -     & $5$ & -           &                                                -        & $5$    & $4320 $      & $\sim0$                                                                \\ \cline{2-10} 
                                              & DSP48E & -          & $21$  & -     & -           & -        & $21$     & $6840$      & $\sim0$          \\ \cline{2-10} 
                                                & FF     & $0$          & $1171$   & -  & -           & $9438$     & $10609$  & $2364480$   & $\sim0$          \\ \cline{2-10} 
                                                & LUT    & $121303$     & $1089$ & -    & $2256$        & -        & $124648$ & $1182240$   & $11$               \\ \hline
\end{tabular}
\label{tab:utilization_real_world}
\end{table*}

\begin{figure*}[t]
	\centering
		\includegraphics[width=6.6in]{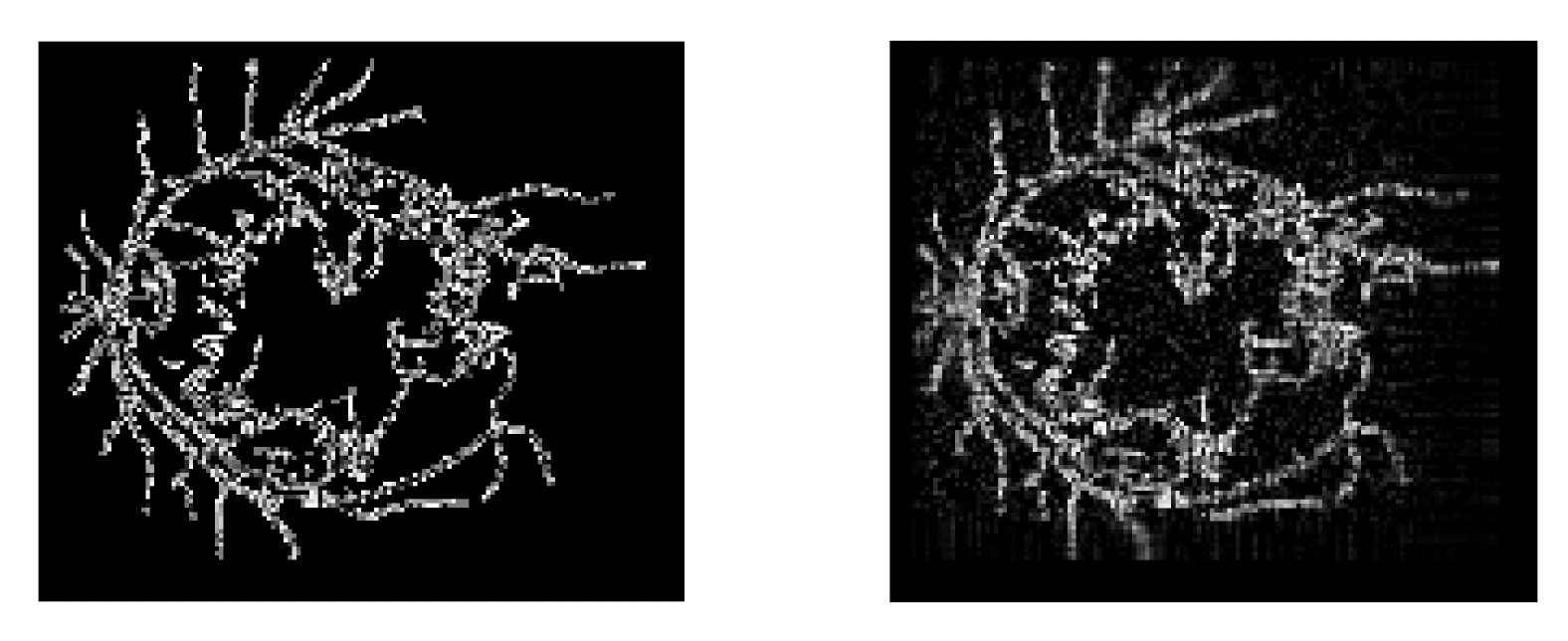} 
\caption{Left: the original retinal angiogram. Right: the approximated image by our sparse Tucker decomposition on the FPGA/CPU hybrid platform.}
	\label{fig:sparse_image}
\end{figure*}

\subsection{Real-World Datasets}
Finally, we verify our accelerator on four real-world sparse tensor data sets~\cite{mcauley2013, BB15,Brent70,carlson2010toward}. In addition, we compare the performance of our accelerator with sparse Tucker decomposition on CPU and with the dense FPGA accelerator in~\cite{zhang2019tucker}. Table~\ref{tab:realExample} shows the detailed run-time and energy consumption of different methods on these datasets.  Table~\ref{tab:utilization_real_world} further shows the overall hardware resource utilization of our method on FPGA. The FPGA design is compiled for each data set in order to achieve the maximum efficiency. We use BRAM\_18K, BDSP48E, FF and LUT to denote block random access memory, digital signal processing elements, flip flops and lookup tables, respectively.

The detailed experiments and results are summarized below:
\begin{itemize}
    \item {\bf Amazon Reviews Datasets~\cite{mcauley2013}.} The modes of this three-way tensor represent users, products, and words, respectively. Each non-zero element in this tensor is the number of times a word appears in a given review. Additionally, we extract one portion of the Amazon reviews tensor of size $20000\times20000\times20000$ and choose the rank of approximation as $R_1=R_2=R_3=32$. We perform 2 power iterations on all modes. The sizes of the tensors and matrices in TTM \eqref{eqn:core_unfolded} are $32 \times 32 \times 20000$ and $32 \times 20000$, respectively. This sparse Tucker factorization involves $9$ calls of QR decomposition on a set of $20000\times 32$ matrices in total to compute the orthogonal factor matrices. Finally, there are totally $8,820$ calls of Kronecker products, which depends on the number of non-zero tensor entries. On this dataset, our hybrid FPGA/CPU platform achieves $1.15\times$ speedup than CPU with only $13.5\%$ energy consumption. Our method also achieves $1091\times$ speedup than the dense Tucker FPGA accelerator~\cite{zhang2019tucker}. 
    
    \item {\bf NELL-2 Datasets~\cite{carlson2010toward}.} This data set is extracted from the Never Ending Language Learner knowledge base. The non-zero entries represent some entity-relation-entity tuples. We extract one portion of the NELL-2 data set and obtain a sparse tensor of size $1000\times1000\times1000$. In addition, we choose our rank of approximation as $R_1=R_2=R_3=16$. We perform 5 power iterations on all modes. The sizes of the tensors and matrices in TTM \eqref{eqn:core_unfolded} are $16 \times 16 \times 1000$ and $16 \times 1000$, respectively. This sparse Tucker factorization involves $15$ calls of QR decomposition on a set of $1000\times 256$ matrices in total to compute the orthogonal factor matrices. Finally, there are totally $432,555$ calls of Kronecker products, which depends on the number of non-zero tensor entries. Our hybrid FPGA/CPU platform achieves $18\times$ speedup and $94.8\%$ energy saving compared with CPU. Our method is also $23.6\times$ faster than the dense FPGA accelerator~\cite{zhang2019tucker}.
    
\item    {\bf Binary $3$-Way Tensor for Parallel Matrix Multiplication~\cite{BB15,Brent70}.} This binary tensor describes the parallel computation process of matrix multiplications. Given two matrices $\mathbf{A}\in{\mathbb{R}^{M \times K}}$ and $\mathbf{B}\in{\mathbb{R}^{K \times N}}$, their product results in a matrix $\mathbf{C}\in{\mathbb{R}^{M \times N}}$. Let $I_1=MK$, $I_2=KN$ and $I_3=MN$, then a binary 3-way tensor $\ten{X}$ can represent the parallel matrix multiplication. The first mode corresponds to the first input matrix $\mat{A}$ with entries in row-major order; the second mode corresponds to the input matrix $\mat{B}$ with entries in row-major order; the third mode corresponds to the output matrix $\mat{C}$ with entries in column-major order. A nonzero entry $x_{i_1 i_2 i_3}=1$ corresponds to a scalar multiplication within the classical matrix multiplication algorithm: the $i_1$-th entry of $\mat{A}$ is multiplied with the $i_2$-th entry of $\mat{B}$, and the result is accumulated into the $i_3$-th entry of $\mat{C}$. The number of nonzero elements in $\ten{X}$ is $nnz=MKN$. We consider the case $M=N=K=5$, which results in a binary tensor $\ten{X}$ with size $25\times 25\times 25$ and a sparsity of $8\times 10^{-3}$. To perform sparse Tucker decomposition on this 3-way binary tensor, we choose an approximation rank of $R_1=R_2=R_3=5$. We perform three steps of high-order power iterations on all modes, leading to 3 TTM in \eqref{eqn:core_unfolded} and totally 6 calls for QR decomposition with column pivoting. Finally, the number of Kronecker products used in this data set is $1,125$. Our meethod achieves $37\times$ and $1.52\times$ speedup than CPU and than the dense FPGA accelerator~\cite{zhang2019tucker}, respectively. Compared with the sparse Tucker decomposition on CPU, our accelerator saves $97.1\%$ energy. 

\item {\bf Retinal Angiogram.} Angiogramy is a medical diagnoictic test that uses X-ray to take picture of the blood vessels. The images, angiogram, are always very sparse. Fig. 6 shows the retinal angiogram of a patient on the left. The size of the original retinal angiogram is $130 \times 150$~\cite{hoover2000locating}. Tucker factorization can also be employed to compress 2-D data, because a matrix is the special case of a tensor. Different from SVD compression of a matrix where the rank is a scalar, a Tucker decomposition allows one to set two rank parameters. We perform a sparse Tucker decomposition with rank
R = [30, 35] on this image. We performed $12$ steps of high-order power iterations on all modes, leading to $12$ TTM in \eqref{eqn:core_unfolded} and totally 24 calls for QR decomposition with column pivoting. We do not need any Kronecker products since the order of the tensor is $2$. Our proposed method achieves $19\times$ speedup than CPU and $1.91\times$ speedup than dense FPGA accelerator~\cite{zhang2019tucker}, and it saves $94.4\%$ energy compared with the sparse Tucker factorization on CPU. Fig.~\ref{fig:sparse_image} compares the original retinal angiogram and the resulting compressed image from our FPGA/CPU hybrid accelerator. The compression ratio is $18.57 \times$. While the image is highly compressed, the essential features, such as blood vessels, are still clearly preserved.  

\end{itemize}

\section{Conclusion}
\label{sec:conclusion}
This paper has proposed a hybrid FPGA-CPU accelerator for sparse Tucker decomposition. On the algorithm level, the Kronecker products have exploited the data sparsity and has significantly reduced the computational complexity. The QR with pivoting method have dramatically reduced the complexity of obtaining the orthogonal mode-n matrix factors. The FPGA modules for the tensor-times-matrix and for the Kronecker products have achieved  $93.519\%$ to $99.514 \%$ energy saving compared with CPU on synthetic benchmarks. The proposed hybrid FPGA-CPU accelerator has been evaluated with both synthetic and realistic sparse tensor data sets. It has achieved $27\times$$\sim$$853\times$ speedup over CPU and $1.167\times$$\sim $$126\times$ speedup over the recently developed dense Tucker FPGA accelerator~\cite{zhang2019tucker} on the synthetic datasets. Our proposed methods have also achieved $1.15\times $$\sim$$1091\times$ speedup and over $95\%$ energy savings on the tested real-world tensor datasets. Our proposed accelerator have significantly outperformed CPU and dense Tucker FPGA accelerator~\cite{zhang2019tucker} when the tensor is very large and sparse.







\bibliographystyle{IEEEtran}
\bibliography{RefList}

\begin{thebibliography}{10}
\providecommand{\url}[1]{#1}
\csname url@samestyle\endcsname
\providecommand{\newblock}{\relax}
\providecommand{\bibinfo}[2]{#2}
\providecommand{\BIBentrySTDinterwordspacing}{\spaceskip=0pt\relax}
\providecommand{\BIBentryALTinterwordstretchfactor}{4}
\providecommand{\BIBentryALTinterwordspacing}{\spaceskip=\fontdimen2\font plus
\BIBentryALTinterwordstretchfactor\fontdimen3\font minus
  \fontdimen4\font\relax}
\providecommand{\BIBforeignlanguage}[2]{{%
\expandafter\ifx\csname l@#1\endcsname\relax
\typeout{** WARNING: IEEEtran.bst: No hyphenation pattern has been}%
\typeout{** loaded for the language `#1'. Using the pattern for}%
\typeout{** the default language instead.}%
\else
\language=\csname l@#1\endcsname
\fi
#2}}
\providecommand{\BIBdecl}{\relax}
\BIBdecl

\bibitem{kolda2009tensor}
T.~G. Kolda and B.~W. Bader, ``Tensor decompositions and applications,''
  \emph{SIAM review}, vol.~51, no.~3, pp. 455--500, 2009.

\bibitem{oseledets2010tt}
I.~Oseledets and E.~Tyrtyshnikov, ``Tt-cross approximation for multidimensional
  arrays,'' \emph{Linear Algebra and its Applications}, vol. 432, no.~1, pp.
  70--88, 2010.

\bibitem{de2000multilinear}
L.~De~Lathauwer, B.~De~Moor, and J.~Vandewalle, ``A multilinear singular value
  decomposition,'' \emph{SIAM journal on Matrix Analysis and Applications},
  vol.~21, no.~4, pp. 1253--1278, 2000.

\bibitem{de2000best}
------, ``On the best rank-1 and rank-($r_1, r_2,\dots, r_n$) approximation of
  higher-order tensors,'' \emph{SIAM journal on Matrix Analysis and
  Applications}, vol.~21, no.~4, pp. 1324--1342, 2000.

\bibitem{candecomp}
J.~D. Carroll and J.~J. Chang, ``Analysis of individual differences in
  multidimensional scaling via an n-way generalization of ``{E}ckart-{Y}oung''
  decomposition,'' \emph{Psychometrika}, vol.~35, no.~3, pp. 283--319, 1970.

\bibitem{harshman1970fpp}
R.~A. Harshman, ``{Foundations of the PARAFAC procedure: Models and conditions
  for an ``explanatory'' multi-modal factor analysis},'' \emph{UCLA Working
  Papers in Phonetics}, vol.~16, no.~1, p.~84, 1970.

\bibitem{vasilescu2002multilinear}
M.~A.~O. Vasilescu and D.~Terzopoulos, ``Multilinear analysis of image
  ensembles: Tensorfaces,'' in \emph{European Conference on Computer
  Vision}.\hskip 1em plus 0.5em minus 0.4em\relax Springer, 2002, pp. 447--460.

\bibitem{vasilescu2003multilinear}
------, ``Multilinear subspace analysis of image ensembles,'' in \emph{IEEE
  Prof. Computer Vision and Pattern Recognition}, vol.~2, 2003, pp. II--93.

\bibitem{vasilescu2005multilinear}
------, ``Multilinear independent components analysis,'' in \emph{IEEE Conf.
  Computer Vision and Pattern Recognition}, vol.~1, 2005, pp. 547--553.

\bibitem{kolda2008scalable}
T.~G. Kolda and J.~Sun, ``Scalable tensor decompositions for multi-aspect data
  mining,'' in \emph{Prof. IEEE Int. Conf. Data Mining}, 2008, pp. 363--372.

\bibitem{batmanghelich2011regularized}
N.~Batmanghelich, A.~Dong, B.~Taskar, and C.~Davatzikos, ``Regularized tensor
  factorization for multi-modality medical image classification,'' in
  \emph{International Conference on Medical Image Computing and
  Computer-Assisted Intervention}.\hskip 1em plus 0.5em minus 0.4em\relax
  Springer, 2011, pp. 17--24.

\bibitem{zhang2016big}
Z.~Zhang, T.-W. Weng, and L.~Daniel, ``Big-data tensor recovery for
  high-dimensional uncertainty quantification of process variations,''
  \emph{IEEE Transactions on Components, Packaging and Manufacturing
  Technology}, vol.~7, no.~5, pp. 687--697, 2016.

\bibitem{zhang2014enabling}
Z.~Zhang, X.~Yang, I.~V. Oseledets, G.~E. Karniadakis, and L.~Daniel,
  ``Enabling high-dimensional hierarchical uncertainty quantification by anova
  and tensor-train decomposition,'' \emph{IEEE Trans. CAD of Integrated
  Circuits and Systems}, vol.~34, no.~1, pp. 63--76, 2014.

\bibitem{zhang2016tensor}
Z.~Zhang, K.~Batselier, H.~Liu, L.~Daniel, and N.~Wong, ``Tensor computation: A
  new framework for high-dimensional problems in eda,'' \emph{IEEE Transactions
  on Computer-Aided Design of Integrated Circuits and Systems}, vol.~36, no.~4,
  pp. 521--536, 2016.

\bibitem{luan2019prediction}
J.~Luan and Z.~Zhang, ``Prediction of multi-dimensional spatial variation data
  via bayesian tensor completion,'' \emph{IEEE Transactions on Computer-Aided
  Design of Integrated Circuits and Systems}, 2019.

\bibitem{novikov2015tensorizing}
A.~Novikov, D.~Podoprikhin, A.~Osokin, and D.~P. Vetrov, ``Tensorizing neural
  networks,'' in \emph{Advances in Neural Information Processing Systems},
  2015, pp. 442--450.

\bibitem{ttrnn2017icml}
Y.~Yang, D.~Krompass, and V.~Tresp, ``Tensor-train recurrent neural networks
  for video classification,'' in \emph{Proc. Int. Conf. Machine Learning},
  vol.~70, 06--11 Aug 2017, pp. 3891--3900.

\bibitem{hawkins2019bayesian}
C.~Hawkins and Z.~Zhang, ``Bayesian tensorized neural networks with automatic
  rank selection,'' \emph{arXiv preprint arXiv:1905.10478}, 2019.

\bibitem{kaya2016high}
O.~Kaya and B.~U{\c{c}}ar, ``High performance parallel algorithms for the
  tucker decomposition of sparse tensors,'' in \emph{2016 45th International
  Conference on Parallel Processing (ICPP)}.\hskip 1em plus 0.5em minus
  0.4em\relax IEEE, 2016, pp. 103--112.

\bibitem{smith2017sparse}
S.~Smith, J.~Park, and G.~Karypis, ``Sparse tensor factorization on many-core
  processors with high-bandwidth memory,'' in \emph{Proc. IEEE Int. Parallel
  and Distributed Processing Symp}, 2017, pp. 1058--1067.

\bibitem{li2015input}
J.~Li, C.~Battaglino, I.~Perros, J.~Sun, and R.~Vuduc, ``An input-adaptive and
  in-place approach to dense tensor-times-matrix multiply,'' in \emph{Proc.
  Int. Conf. High Performance Computing, Networking, Storage and Analysis},
  2015, pp. 1--12.

\bibitem{kim2015compression}
Y.-D. Kim, E.~Park, S.~Yoo, T.~Choi, L.~Yang, and D.~Shin, ``Compression of
  deep convolutional neural networks for fast and low power mobile
  applications,'' \emph{arXiv preprint arXiv:1511.06530}, 2015.

\bibitem{srivastava2019t2s}
N.~Srivastava, H.~Rong, P.~Barua, G.~Feng, H.~Cao, Z.~Zhang, D.~Albonesi,
  V.~Sarkar, W.~Chen, P.~Petersen \emph{et~al.}, ``T2s-tensor: Productively
  generating high-performance spatial hardware for dense tensor computations,''
  in \emph{International Symp. Field-Programmable Custom Computing Machines
  (FCCM)}, 2019, pp. 181--189.

\bibitem{huang2019high}
W.-P. Huang, B.~P. Kwan, W.~Ding, B.~Min, R.~C. Cheung, L.~Qi, and H.~Yan,
  ``High performance hardware architecture for singular spectrum analysis of
  hankel tensors,'' \emph{Microprocessors and Microsystems}, vol.~64, pp.
  120--127, 2019.

\bibitem{zhang2019tucker}
K.~Zhang, X.~Zhang, and Z.~Zhang, ``Tucker tensor decomposition on fpga,'' in
  \emph{Intl. Conf. Computer Aided Design}, no. 6A.1, 2019.

\bibitem{srivastava2020tensaurus}
N.~Srivastava, H.~Jin, S.~Smith, H.~Rong, D.~Albonesi, and Z.~Zhang,
  ``Tensaurus: A versatile accelerator for mixed sparse-dense tensor
  computations,'' in \emph{2020 IEEE International Symposium on High
  Performance Computer Architecture (HPCA)}.\hskip 1em plus 0.5em minus
  0.4em\relax IEEE, 2020, pp. 689--702.

\bibitem{roohi2016dynamic}
S.~F. Roohi, D.~Zonoobi, A.~A. Kassim, and J.~L. Jaremko, ``Dynamic mri
  reconstruction using low rank plus sparse tensor decomposition,'' in
  \emph{IEEE Int. Conf. Image Processing}, 2016, pp. 1769--1773.

\bibitem{fillard2005extrapolation}
P.~Fillard, V.~Arsigny, X.~Pennec, P.~M. Thompson, and N.~Ayache,
  ``Extrapolation of sparse tensor fields: Application to the modeling of brain
  variability,'' in \emph{Biennial International Conference on Information
  Processing in Medical Imaging}.\hskip 1em plus 0.5em minus 0.4em\relax
  Springer, 2005, pp. 27--38.

\bibitem{van2000ubiquitous}
C.~F. Van~Loan, ``The ubiquitous kronecker product,'' \emph{Journal of
  computational and applied mathematics}, vol. 123, no. 1-2, pp. 85--100, 2000.

\bibitem{golub1996matrix}
G.~H. Golub and C.~F. V.~V. Loan, \emph{Matrix Computations, 3rd ed}.\hskip 1em
  plus 0.5em minus 0.4em\relax Baltimore, MD: Johns Hopkins University Press,
  1996.

\bibitem{golub1971singular}
G.~H. Golub and C.~Reinsch, ``Singular value decomposition and least squares
  solutions,'' in \emph{Linear Algebra}.\hskip 1em plus 0.5em minus 0.4em\relax
  Springer, 1971, pp. 134--151.

\bibitem{gustavson1972some}
F.~G. Gustavson, ``Some basic techniques for solving sparse systems of linear
  equations,'' in \emph{Sparse matrices and their applications}.\hskip 1em plus
  0.5em minus 0.4em\relax Springer, 1972, pp. 41--52.

\bibitem{tew2016investigation}
P.~A. Tew, ``An investigation of sparse tensor formats for tensor libraries,''
  Ph.D. dissertation, Massachusetts Institute of Technology, 2016.

\bibitem{mcauley2013}
J.~McAuley and J.~Leskovec, ``Hidden factors and hidden topics: understanding
  rating dimensions with review text,'' in \emph{Proceedings of the 7th ACM
  conference on Recommender systems}.\hskip 1em plus 0.5em minus 0.4em\relax
  ACM, 2013, pp. 165--172.

\bibitem{BB15}
A.~R. Benson and G.~Ballard, ``A framework for practical parallel fast matrix
  multiplication,'' in \emph{Proc. ACM SIGPLAN Symposium on Principles and
  Practice of Parallel Programming}, 2015, pp. 42--53.

\bibitem{Brent70}
R.~P. Brent, ``Algorithms for matrix multiplication,'' Stanford University,
  Tech. Rep., 1970.

\bibitem{carlson2010toward}
A.~Carlson, J.~Betteridge, B.~Kisiel, B.~Settles, E.~R. Hruschka~Jr., and T.~M.
  Mitchell, ``Toward an architecture for never-ending language learning.'' in
  \emph{AAAI}, vol.~5, 2010, p.~3.

\bibitem{hoover2000locating}
A.~Hoover, V.~Kouznetsova, and M.~Goldbaum, ``Locating blood vessels in retinal
  images by piecewise threshold probing of a matched filter response,''
  \emph{IEEE Transactions on Medical imaging}, vol.~19, no.~3, pp. 203--210,
  2000.

\end{thebibliography}
\begin{IEEEbiography}[{\includegraphics[width=1in,height=1.25in,clip,keepaspectratio]{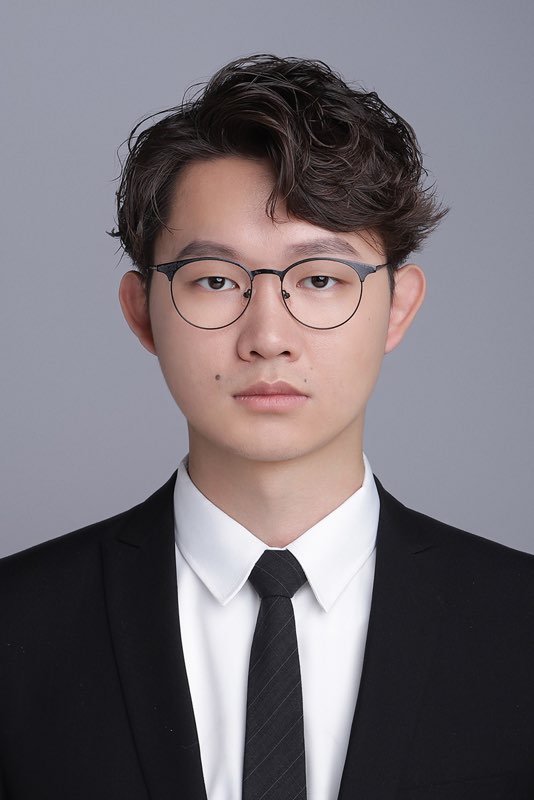}}]{Weiyun Jiang} received his B.Sc. degree in Electrical Engineering from University of California at Santa Barbara, Santa Barbara, CA, in 2020. Currently he is a graduate student of Electrical Engineering at Stanford University, Stanford, CA. His research interests include algorithm/hardware co-design for tensor data analysis and machine learning. 
\end{IEEEbiography}
\begin{IEEEbiography}[{\includegraphics[width=1in,height=1.25in,clip,keepaspectratio]{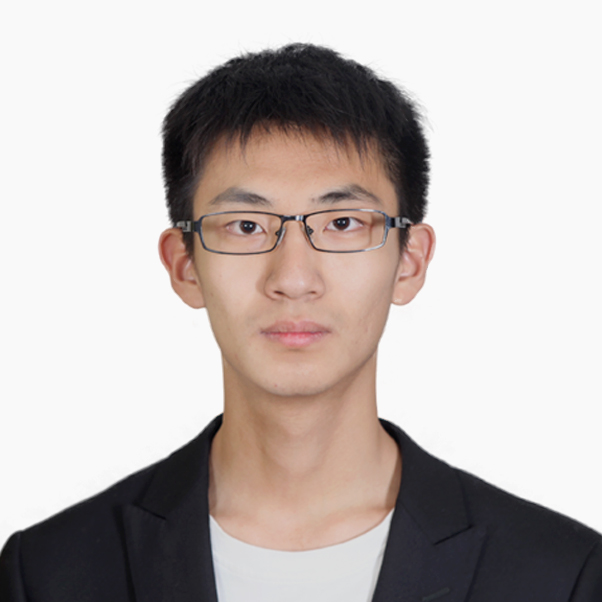}}]{Kaiqi Zhang} received the B.Sc. degree in electronic engineering from Tsinghua University, Beijing, China, in 2016, the M.S. degree in electrical and computer engineering from the University of California at Davis, Davis, CA, in 2018, and he is now pursuing his Ph.D. degree in electrical and computer engineering from the University of California at Santa Barbara, Santa Barbara, CA.
\end{IEEEbiography}

\begin{IEEEbiography}[{\includegraphics[width=1in,height=1.25in,clip,keepaspectratio]{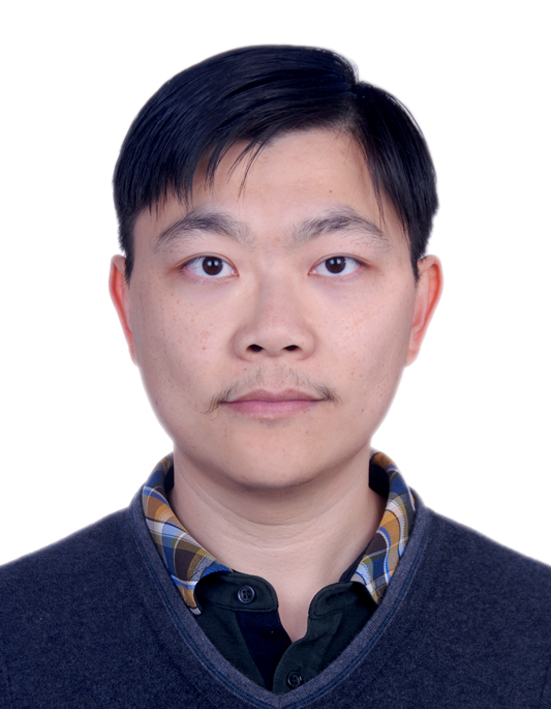}}]{Colin Yu Lin} received his B.Sc. degree in electronic engineering from Sun Yat-Sen University, Guangzhou, China, in 2005, the M.E. degree in computer engineering from the University of Chinese Academy of Sciences, Beijing, China, in 2008,and the Ph.D. degree in electrical and electronic engineering from the University of Hong Kong, Hong Kong, in 2012.
From 2011 to 2012, he was a Visiting Student Researcher with the Department of Electrical Engineering and Computer Sciences and the Berkeley Wireless Research Center, University of California at Berkeley, Berkeley, CA, USA. He was an Assistant Professor with the System on Programmable Chip Research Department, Institute of Electronics, Chinese Academy of Sciences, Beijing, China, from 2012 to 2016. He is currently a Software Development Senior Manager with Data Center Group, Xilinx, Inc. His current research interests include FPGA architecture, CAD for FPGAs, high level synthesis, and FPGA for high performance computing.
\end{IEEEbiography}

\begin{IEEEbiography}[{\includegraphics[width=1in,height=1.25in,clip,keepaspectratio]{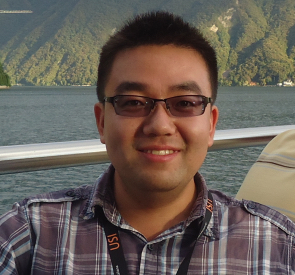}}]{Feng Xing} received his B.Sc degree from Wuhan University, China and M.Sc from University of Lille, French on pure mathematics and applied mathematics. He got his Ph.D. from “Maison de la Simulation”, CEA Saclay on high performance computing in 2014. Then he worked as postdoc researcher at INRIA French and BRGM French for two years on high performance geothermal simulation. He is currently a software development manager at Xilinx, Inc.
\end{IEEEbiography}

\begin{IEEEbiography}[{\includegraphics[width=1in,height=1.25in,clip,keepaspectratio]{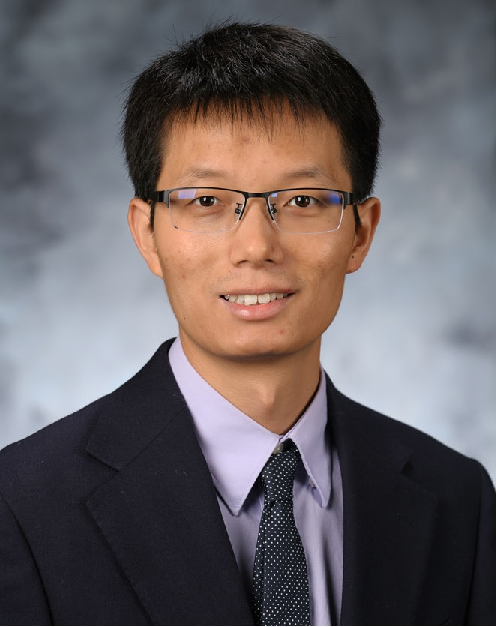}}]{Zheng Zhang} (M'15) has been an Assistant Professor of Electrical and Computer Engineering with the University of California at Santa Barbara (UCSB), since July 2017. He received his Ph.D in Electrical Engineering and Computer Science from Massachusetts Institute of Technology (MIT), Cambridge, MA, in 2015. His research interests include uncertainty quantification and tensor computation. The applications of his research include design automation of nano-scale electronics and photonics, algorithm/hardware co-design of high-dimensional, robust and safe machine learning systems.

Dr. Zhang received three best paper awards from IEEE Transactions: the Best Paper Award of IEEE Transactions on Computer-Aided Design of Integrated Circuits and Systems in 2014, two Best Paper Awards of IEEE Transactions on Components, Packaging and Manufacturing Technology in 2018 and 2020, respectively. He also received three Best Paper Awards at international conferences. His PhD dissertation won the ACM SIGDA Outstanding Ph.D Dissertation Award in Electronic Design Automation in 2016, and the Best Thesis Award from the Microsystems Technology Laboratory of MIT in 2015. He received a NSF CAREER Award in 2019 and a Facebook Research Award in 2020.
\end{IEEEbiography}

\end{document}